\documentclass[prl,showpacs,preprintnumbers,amsmath,aps,twocolumn]{revtex4}

\usepackage{graphicx}
\usepackage{xypic}
\usepackage{amsmath}
\usepackage{amsfonts}
\usepackage{amssymb}
\usepackage{amscd}

\def\duzomniejsze{<\kern-.7mm<}
\def\duzowieksze{>\kern-.7mm>}

\def\textbf#1{{\bf #1}}
\def\be{\begin{equation}}
\def\ee{\end{equation}}
\def\bea{\begin{eqnarray}}
\def\eea{\end{eqnarray}}
\def\bse{\begin{subequations}}
\def\ese{\end{subequations}}
\newcommand{\bei}{\begin{itemize}}
\newcommand{\eei}{\end{itemize}}
\newcommand{\bee}{\begin{enumerate}}
\newcommand{\eee}{\end{enumerate}}

\def\hcal{{\cal H}}
\def\tr{{\rm Tr}}

\def\>{\rangle}
\def\<{\langle}

\def\dt#1{{{\kern -.0mm\rm d}}#1\,}

\newtheorem{definition}{Definition}
\newtheorem{proposition}{Proposition}
\begin{document}

\title{An additive and operational entanglement measure: conditional entanglement of mutual information}
\author{Dong Yang$^{1, 2}$}
\author{Micha\l{} Horodecki$^{3,4}$}
\author{Z. D. Wang$^{1}$}

\affiliation{$^{1}$Department of Physics and Center of Theoretical and Computational Physics, The University of
Hong Kong, Pokfulam Road, Hong Kong, China\\
$^{2}$Laboratory for Quantum Information and College of Mechatronics Engineering,
China Jiliang University, Hangzhou, Zhejiang 310018, China\\
 $^{3}$Institute of Theoretical Physics and
Astrophysics, University of Gda\'nsk, 80--952 Gda\'nsk, Poland \\
 $^{4}$National Quantum Information Centre of Gda\'nsk, University of Gda\'nsk, 80--952 Sopot, Poland}

\date{\today}

\begin{abstract}
Based on the monogamy of entanglement, we develop the technique of quantum conditioning to build an {\it additive} entanglement measure: the conditional entanglement of mutual information. Its {\it operational} meaning is elaborated to be the minimal net "flow of qubits" in the process of partial state merging. The result and conclusion can also be generalized to multipartite entanglement cases.
\end{abstract}


\maketitle
Entanglement, as a key resource and ingredient in quantum information and computation as well as communication, plays a crucial role in quantum information theory. It is necessary to quantify entanglement from different standpoints. A number of entanglement measures have been proposed, and their properties have been explored extensively (see, e.g., Ref.\cite{4H,PV} and references therein). Nevertheless several questions are needed to be answered, especially: i) How to systematically introduce new entanglement measures. It is commonly accepted that an appropriate entanglement measure is necessarily non-increasing under local operations and classical communication (LOCC). But few approaches to construct entanglement measures are known. For example, the entanglement of formation $E_f$ \cite{Bennett1} is established via the technique of "convex roof" and the relative entropy of entanglement $E_r$ \cite{Vedral} is based on a concept of "distance". ii) The operational meaning. Entanglement measures are largely studied by the monotonicity under LOCC operations, but little is known for the operational meaning except the distillable entanglement $E_d$ \cite{Bennett1} and entanglement of cost $E_c$ \cite{Hayden}. Just recently, a new paradigm to explain entanglement measures is proposed based on quantum communication \cite{Oppenheim}, where squashed entanglement $E_{sq}$ \cite{Christandl} obtains its meaning. iii) Additivity. It is a very desirable property that can largely reduce computation of entanglement. Since quantum mechanics is statistical, often operational meaning of entanglement measures is acquired only in asymptotic regime of many copies of given state. For additive measures, it is reduced to a single copy. Additivity holds for squashed entanglement $E_{sq}$ \cite{Christandl} and logarithmic negativity $E_N $ \cite{ZyczkowskiHSP-vol,Vidal}, and is conjectured to hold for $E_f$, but $E_r$ is nonadditive \cite{VW}. iv) Multipartite entanglement. It is more difficult to design multipartite entanglement measures, hence it would be good, if a bipartite one can be easily extended to multipartite regime.

In this paper, based on the monogamy of entanglement, we develop the technique of quantum conditioning of correlation function to construct entanglement measures. Taking the quantum mutual information as the correlation function, we formulate a new entanglement measure---the conditional entanglement of mutual information. Remarkably, it is {\it additive} with an {\it operational} meaning and can straightforwardly be generalized to multipartite cases.

Let us begin with the question how to build an entanglement measure. The monogamy of entanglement \cite{Werner} is a good starting point. It tells that entanglement is a type of quantum correlation that cannot be shared. This feature is distinct from the classical correlation that can be shared. A simple example is the Bell state $|\Phi\>_{AB}=1/\sqrt{2}(|00\>+|11\>$ between Alice and Bob. Monogamy of the pure entangled state $|\Phi\>_{AB}$ excludes the possibility that any other party could correlate with. It is different for the classical correlated state $\rho_{AB}=1/2(|00\>\<00|+|11\>\<11|)$. Obviously another party Charlie can share the correlation with the form $\rho_{ABC}=1/2(|000\>\<000|+|111\>\<111|)$. The example is the extremal case in which quantum correlation and classical one are well separated. However it is not the case for a generic mixed state. A correlation function $f(A:B)$ \cite{note,HendersonVedral}, for instance quantum mutual information, usually contains quantum correlation and classical one, and is 'dirty' in the sense that quantum correlation and classical one are interwound in a complex way that cannot be separated neatly. How can we 'distill' a 'neat' quantum correlation? The technique is quantum extension and quantum conditioning. Quantum extension means that given a state $\rho_{AB}$, we embed it into a larger state $\rho_{AA'BB'}$ such that $\rho_{AB}$ is the reduced state of $\rho_{AA'BB'}$, i.e. $tr_{A'B'}\rho_{AA'BB'}=\rho_{AB}$. Apparently $f(AA':BB')$ is larger than $f(A:B)$. To return a correlation measure for $\rho_{AB}$, we consider difference $f(AA':BB')-f(A':B')$. Now, let us imagine for a while that quantum (q) and classical (c)  correlations sum up in a simple way. Then due to unsharability of q we can write $f(AA':BB')=q(AB)+q(A'B') +c(AA':BB')\equiv q_1 +q_2 + c_{12}$, while $f(A':B')=q(A':B') + c(A':B')\equiv q_2 + c_2$. Subtracting we get $q_1$ (i.e. what we want) plus the difference $c_{12}-c_2$ which, as we have seen, can be zero because classical correlations are sharable.  In general it will not vanish, so we take infimum over extensions, trying to squash out the classical correlations as much as we can. The infimum of the difference must be of purely quantum origin, hence we treat it as a correction to our initial, oversimplified assumption.

For given function $f(\cdot)$ quantifying correlation, we have two candidates for its conditioned version \bse \bea
C_f^s(\rho_{AB})&=&\inf[f(\rho_{AA':BB'})-f(\rho_{A':B'})],\\
C_f^a(\rho_{AB})&=&\inf[f(\rho_{A:BE})-f(\rho_{A:E})], \eea \ese where infimum is taken over all extensions $\rho_{AA'BB'}$  ($\rho_{ABE}$) of $\rho_{AB}$. $C_f^s(\cdot)$ is the symmetric conditioned version of $f$ while $C_f^a(\cdot)$ the asymmetric one. Note that the above definition is similar to that of conditional entropy \cite{HOW} $S(A|B)=S(AB)-S(B)$ with $S(\rho)$ as the von Neumann entropy $S(\rho)=-\tr\rho\log{\rho}$, and thus referred to as {\it conditional entanglement}. As a matter of fact, squashed entanglement can be constructed by taking asymmetric conditioning of mutual information, $E_{sq}(\rho_{AB})={1\over 2}\inf\{I(A:BE)-I(A:E)\}\equiv {1\over 2}\inf I(A:B|E)$, where $I(X:Y)=S(X)+S(Y)-S(XY)$ is quantum mutual information and $I(A:B|E)=S(AE)+S(BE)-S(ABE)-S(E)$ is conditional mutual information. It is notable that $I(A:BE)-I(A:E)=I(AE:B)-I(E:B)$ is symmetric w.r.t. systems $AB$ though each term in the formula is asymmetric w.r.t. both parties. This gives the possibility to build symmetric entanglement measures by asymmetric conditioning. It is surprising that a 'neat' quantum correlation can be obtained by subtracting two 'dirty' functions. Does this approach really work? The answer is YES (see Ref.\cite{Web} to  systematically introduce new entanglement measures based on quantum conditioning). We illustrate that a new entanglement measure can indeed be constructed by taking $f$ to be quantum mutual information in the symmetric version. We add a factor $1/2$ and denote it by $E_I$. Most intriguingly, we show below that $E_I$ is additive, has an operational meaning, and can be directly generalized to multipartite states where the factor $1/2$ has a good reason to exist. {\definition ~} Let $\rho_{AB}$ be a mixed state on a bipartite Hilbert space ${\cal H}_{A}\otimes {\cal H}_{B}$. The conditional entanglement of mutual information for $\rho_{AB}$ is defined as \be E_{I}(\rho_{AB})=\inf{1\over 2}\{I(AA':BB')-I(A':B')\}, \ee
 where the infimum is taken over all extensions of $\rho_{AB}$, i.e., over all states satisfying the equation $\tr_{A'B'}\rho_{AA'BB'}=\rho_{AB}$.

To justify that $E_{I}$ is an appropriate entanglement measure, we now elaborate that it does satisfy two essential axioms that an entanglement measure should obey \cite{4H}.

{\it 1. Entanglement does not increase under local operations and classical communication (LOCC) i. e. $E_I(\Lambda(\rho))\le E_I(\rho)$, for any LOCC operation $\Lambda$.} The monotonicity under LOCC implies that entanglement remains invariant under local unitary transformations. This comes from the fact local unitary transformations are reversible LOCC. The convexity of entanglement used to be considered as a mandatory ingredient of the mathematical formulation of monotonicity \cite{4H,Plenio}. At present,  the convexity is thought to be merely a convenient mathematical property. Also there is a common agreement that the strong monotonicity---monotonicity {\it on average} under LOCC is unnecessary but useful \cite{4H,Plenio}. Many known existing entanglement measures are convex and satisfy the strong monotonicity. We will show that $E_I$ satisfies the strong monotonicity.

Since, as we will see further, $E_I$ is convex, it is sufficient to prove  that  $E_{I}$ is non-increasing under a local measurement  \cite{Vidal-mon2000} (w.l.o.g we can check it only on Alice side) namely, $E_{I}(\rho_{AB})\ge \sum_k p_kE_{I}(\tilde{\rho}_{AB}^{k}),$ where $\tilde{\rho}_{AB}^{k}=A_{k}\rho_{AB}A_{k}^{\dagger}/p_k$, $p_k=trA_{k}\rho_{AB}A_{k}^{\dagger}$, and $\sum_kA_{k}^{\dagger}A_{k}=I_A$. Another way to describe the measurement process is as following. First, one attaches two ancillary systems $A_0$ and $A_1$ in states $|0\>_{A_0}$ and $|0\>_{A_1}$ to system $AB$. Secondly, a unitary operation $U_{AA_0A_1}$ on $AA_0A_1$ is performed. Thirdly, the system $A_1$ is traced out to get the state as $ \tilde{\rho}_{A_0AB}=\sum_{k}A_{k}\rho_{AB}A_{k}^{\dagger}\otimes (|k\>\<k|)_{A_0}. $ Now for any extension state $\rho_{AA'BB'}$, we get the state after the measurement on A, $ \tilde{\rho}_{A_0AA'BB'}=\sum_{k}A_{k}\rho_{AA'BB'}A_{k}^{\dagger}\otimes (|k\>\<k|)_{A_0} =\sum_{k}p_k\tilde{\rho}_{AA'BB'}^{k}\otimes (|k\>\<k|)_{A_0}\nonumber.$  Most crucially, we have
 \bse \bea
&&I(\rho_{AA':BB'})-I(\rho_{A':B'})\nonumber\\
&=&I(0_{A_0A_1}\otimes\rho_{AA':BB'})-I(\rho_{A':B'})\\
&=&I(U_{A_0A_1A}(0_{A_0A_1}\otimes\rho_{AA':BB'}))-I(\rho_{A':B'})\label{se:non1}\\
&\ge& I(\tilde{\rho}_{A_0AA':BB'})-I(\tilde{\rho}_{A':B'})\label{se:non2}\\
&=&\sum_{k}p_k[I(\tilde{\rho}_{AA':BB'}^{k})-I(\tilde{\rho}_{A':B'}^{k})]\nonumber\\
&+&\sum_kp_k I(\tilde{\rho}_{A':B'}^{k})-I(\tilde{\rho}_{A':B'})\nonumber\\
&+&S(\tilde{\rho}_{BB'})-\sum_{k}p_kS(\tilde{\rho}_{BB'}^{k})\nonumber\\
&=&\sum_{k}p_k[I(\tilde{\rho}_{AA':BB'}^{k})-I(\tilde{\rho}_{A':B'}^{k})]\nonumber\\
&+&\chi(BB')+\chi(A'B')-\chi(A')-\chi(B')\nonumber\\
&\ge&\sum_{k}p_k[I(\tilde{\rho}_{AA':BB'}^{k})-I(\tilde{\rho}_{A':B'}^{k})]\label{se:non3} \eea \ese where $\chi(\rho)=S(\rho)-\sum_kp_kS(\rho^{k})$ is the Holevo quantity of the ensemble $\{p_k,\rho^{k}\}$. The equality of (\ref{se:non1}) follows from invariance of quantum mutual information  under local unitary operations, while the inequalities of (\ref{se:non2}) and (\ref{se:non3}) stem from, respectively, the facts that quantum mutual information and the Holevo quantity are non-increasing under tracing out a subsystem. Consequently, we have proved that $E_{I}$ is non-increasing on average under LOCC operations.

{\it 2. Entanglement is not negative and is zero for separable states.} The inequality $E_{I}\ge 0$ comes from the fact that the quantum mutual information is non-increasing under tracing subsystems of both sides. For a separable state $\rho_{AB}$, it can always be decomposed into a separable form: $\rho_{AB}=\sum_{i,j}p_{ij}\phi_{A}^{i}\otimes \phi_{B}^{j}$. An extension state may be chosen to be $ \rho_{AA'BB'}=\sum_{i,j}p_{ij}\phi_{A}^{i}\otimes(|i\>\<i|)_{A'}\otimes\phi_{B}^{j}\otimes(|j\>\<j|)_{B'}. $ It is obvious that $I(AA':BB')=I(A':B')$, and thus $E_{I}=0$ for separable states.

{\it Continuity.} The conditional entanglement of mutual information is asymptotically continuous, i.e. if $|\rho_{AB}-\sigma_{AB}|\leq \epsilon$, then $|E_I(\rho)-E_I(\sigma)|\leq K \epsilon \log d + O(\epsilon)$, where $|\cdot|$ is the trace norm for matrix, $K$ is a constant, $d=\dim{\hcal_{AB}}$, and $O(\epsilon)$ is a function that depends only on $\epsilon$ (in particular, it does not depend on dimension) and satisfies $\lim_{\epsilon\to 0}O(\epsilon)=0$.

The proof of the asymptotic continuity is similar to that for squashed entanglement and is presented in the Appendix of Ref.\cite{Web}.

{\it Convexity.} $E_{I}$ is convex, i.e., $E_{I}(\lambda \rho+(1-\lambda)\sigma)\le \lambda E_{I}(\rho)+(1-\lambda)E_{I}(\sigma)$ for $0\le \lambda \le 1$.

{\it Proof ~}For any extension states $\rho_{AA'BB'}$ and $\sigma_{AA'BB'}$, we consider the extension state $ \tau_{AA'A''BB'B''}=\lambda \rho_{AA'BB'}\otimes(|0\>\<0|)_{A''}\otimes(|0\>\<0|)_{B''} +(1-\lambda)\sigma_{AA'BB'}\otimes(|1\>\<1|)_{A"} \otimes(|1\>\<1|)_{B''} $, and have $ I(\tau_{AA'A'':BB'B''})-I(\tau_{A'A'':B'B''}) =\lambda [I(\rho_{AA':BB'})-I(\rho_{A':B'})] +(1-\lambda) [I(\sigma_{AA':BB'})- I(\sigma_{A':B'})]. $ This implies $E_{I}$ is convex.

An immediate corollary of convexity is that $E_{I}\le E_f$ and furthermore $E_{I}\le E_c$ due to the following additivity.

{\proposition ~} $ E_{I}(\rho_{AB}\otimes\sigma_{CD})=E_{I}(\rho_{AB})+E_{I}(\sigma_{CD}). $

{\it Proof ~} On the one hand, for any extension states $\rho_{AA'BB'}$ and $\sigma_{CC'DD'}$, $\rho_{AA'BB'}\otimes \sigma_{CC'DD'}$ is an extension state of $\rho_{AB}\otimes\sigma_{CD}$.\bea
&~&I(AA'CC':BB'DD')-I(A'C':B'D')\nonumber\\
&=&I(AA':BB')-I(A':B')\nonumber\\
&+&I(CC':DD')-I(C':D'). \eea So $E_{I}(\rho_{AB}\otimes\sigma_{CD})\le E_{I}(\rho_{AB})+E_{I}(\sigma_{CD})$ holds.

On the other hand,  for extension states $\tau_{ACE':BDF'}$ of $\rho_{AB}\otimes\sigma_{CD}$, $\tau_{ACE':BDF'}$ is an extension state of $\rho_{AB}$ and $\tau_{CE':DF'}$ is an extension state of $\sigma_{CD}$. Therefore we have \bea
&~&I(ACE':BDF')-I(E':F')\nonumber\\
&=&I(ACE':BDF')-I(CE':DF')\nonumber\\
&+&I(CE':DF')-I(E':F'). \eea This means that $E_{I}(\rho_{AB}\otimes\sigma_{CD})\ge E_{I}(\rho_{AB})+E_{I}(\sigma_{CD})$. So we have finally the additivity equality.

It is quite remarkable that the  property of additivity is rather easy to prove for conditional entanglement while it is extremely tough for other candidates. The reason lies in that the conditional entanglement is naturally  super-additive while others are usually sub-additive.

Before we elaborate on the operational meaning of the measure $E_I$, we briefly recall that of quantum conditional mutual information \cite{Devetak}, in which the quantum mutual information one \cite{Abeyesinghe} corresponds to a special case. Quantum conditional mutual information is given the operational meaning in the process of quantum state redistribution \cite{Devetak}. The situation is depicted in FIG 1: Initially $XY$ is with Alice, $Z$ with Bob. $R$ is the reference system such that $\Phi_{RXYZ}$ is pure. The task is that Alice sends $Y$ to Bob while the final state is still in the pure state $\Phi_{RXYZ}$. Alice and Bob share entanglement for free and have an ideal quantum channel to communicate. No classical communication is allowed. To accomplish the task, the minimal amount of qubits that are required to transfer from Alice to Bob is $Q=1/2I(R:Y|Z)$.
\begin{figure}
\label{fig:redis} \centering
\includegraphics[scale=0.5]{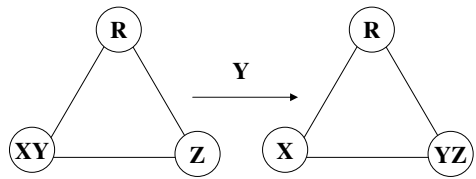}
\caption{Quantum state redistribution}
\end{figure}

In a recent paper \cite{Oppenheim}, the squashed entanglement received the operational meaning with the aid of that of conditional mutual information. It gives a hint for finding the operational meaning for $E_I$ since it can be regarded as a measure constructed in the same spirit. Does the conditional function $\frac{1}{2}\{I(AA':BB')-I(A':B')\}$ have an operational meaning? It turns out that it does. Even more, it is a  conservative quantity, which describes a process, but depends only on the initial and final state. The scenario where it works is a process called partial state merging (PSM). Here we take the name---partial state merging that is somewhat different from the original one in \cite{HOW}. The situation of PSM is depicted in FIG 2: Initially $AA'$ is with Alice and $BB'$ with Bob, $E$ is with the merging center, and the whole state $\Phi_{AA'BB'E}$ is pure. The task is to transfer $A$ and $B$ to the center while the final state remains the same. There is infinite entanglement and an ideal quantum channel between Alice (Bob) and the center. But no entanglement and no channel exists between Alice and Bob. No classical communication is allowed between Alice (Bob) and the center. To accomplish the task, the minimal net flow of qubits to the center is none other than $Q=\frac{1}{2}\{I(AA':BB')-I(A':B')\}$, where the flow into the center is regarded as positive flow while that out is negative one. There are many different routes to merge $A$ and $B$. Dramatically, the net flow is a conservative quantity independent of the different routes of merging. Without loss of generality, we take the two typical routes in FIG 3 to show this. In the routes $I$ and $II$, the net flow of qubits to $E$ is calculated as \bea
Q_I&=&1/2\{I(BB':A|E)+I(A':B|EA)\}\nonumber\\
&=&1/2\{I(AA':BB')-I(A':B')\},\nonumber\\
Q_{II}&=&1/2\{I(BB':AA')+0-I(A':B')\},\nonumber \eea where the relation $S(X)=S(Y)$ is used when $XY$ is in a pure state. Of course there are other routes, however the net flow to the center is the same.
\begin{figure}
\label{fig:merge} \centering
\includegraphics[scale=0.4]{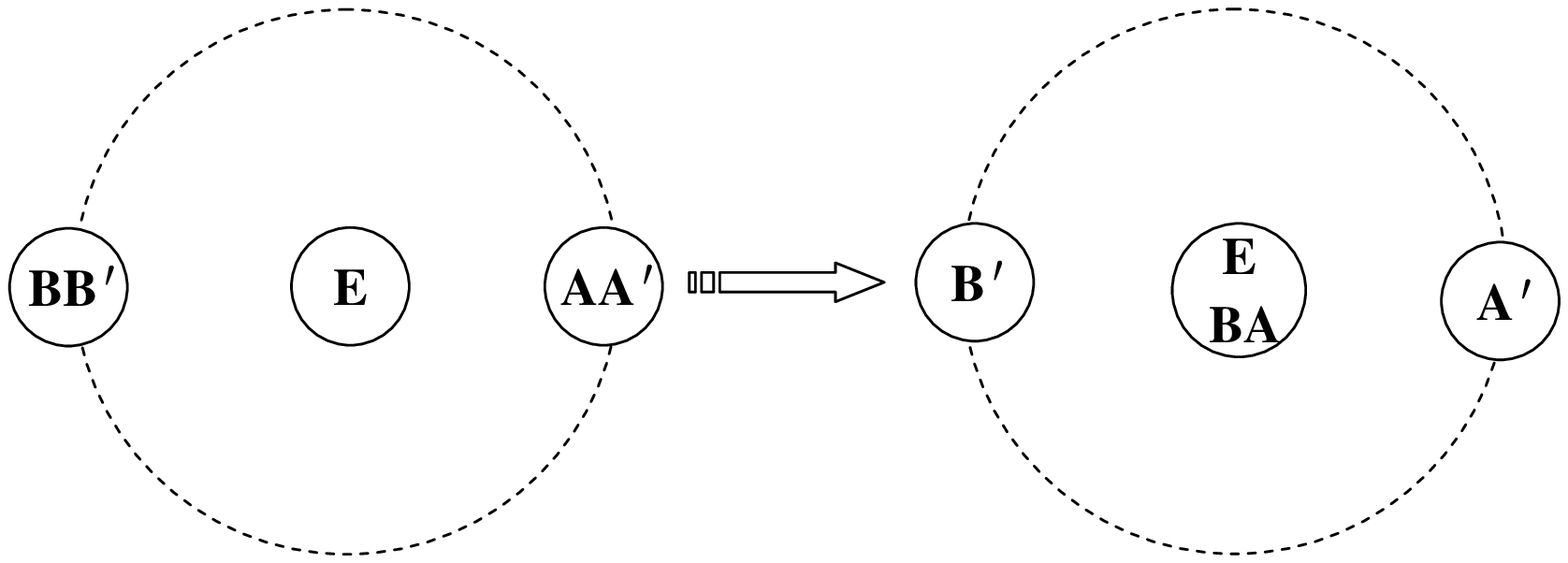}
\caption{Partial state merging}
\end{figure}

\begin{figure}
\label{fig:route} \centering
\includegraphics[scale=0.5]{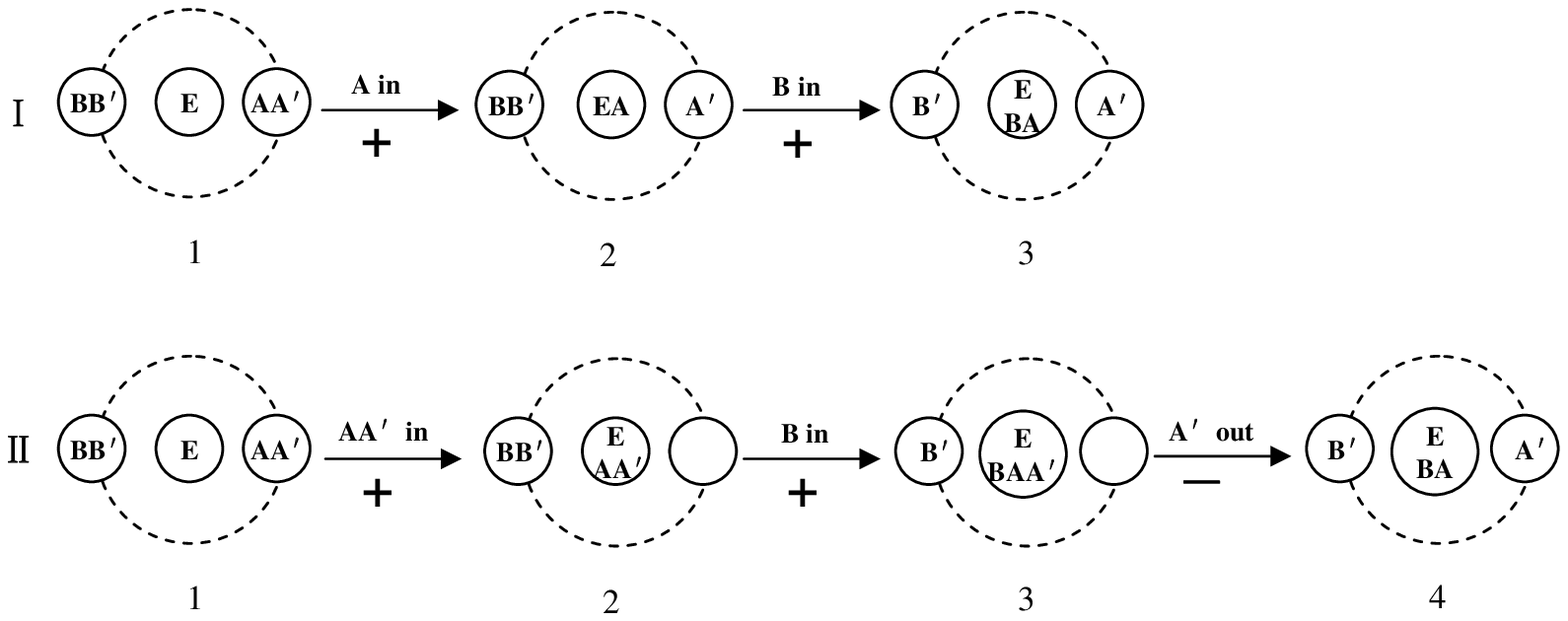}
\caption{Two typical routes}
\end{figure}
Given the operational meaning of the quantity $Q=\frac{1}{2}\{I(AA':BB')-I(A':B')\}$, we immediately obtain the operational meaning of $E_I$. {\proposition ~} For a given mixed state $\rho_{AB}$ to be merged to a center, the conditional entanglement of mutual information is the minimal net flow of qubits to the center with the optimal side-information $\rho_{A'B'}$.

Notice that for separable state $\rho_{AB}$, there always exist the side-information $\rho_{A'B'}$ such that the net flow of qubits to the center is zero. The more entangled $\rho_{AB}$ is, the greater  is the flow of qubits to the merging center.

The result and conclusion can be straightforwardly generalized to the multipartite case where the multipartite version of $E_I$ is defined as $ E_{I}=\inf\frac{1}{2}\{I_n(A_1A'_1:\cdots:A_nA'_n)-I_n(A'_1:\cdots:A'_n)\}$, and $I_n=\sum_iS(A_i)-S(A_1\cdots A_n)$ is the multipartite mutual information \cite{RH}.

{\proposition ~} The conditional entanglement for multipartite mutual information is additive, \be E_{I}(\rho_{A_1\cdots A_n}\otimes\sigma_{B_1\cdots B_n})
=E_{I}(\rho_{A_1\cdots A_n})+E_{I}(\sigma_{B_1\cdots B_n}).\nonumber\\
\ee {\proposition ~} For a multipartite mixed state $\rho_{A_1\cdots A_n}$ to be merged to a center, the conditional entanglement of mutual information is the minimal net flow of qubits to the center with the optimal side-information $\rho_{A_{1}^{'}\cdots A_{n}^{'}}$.

\begin{figure}
\label{fig:psdm} \centering
\includegraphics[scale=0.43]{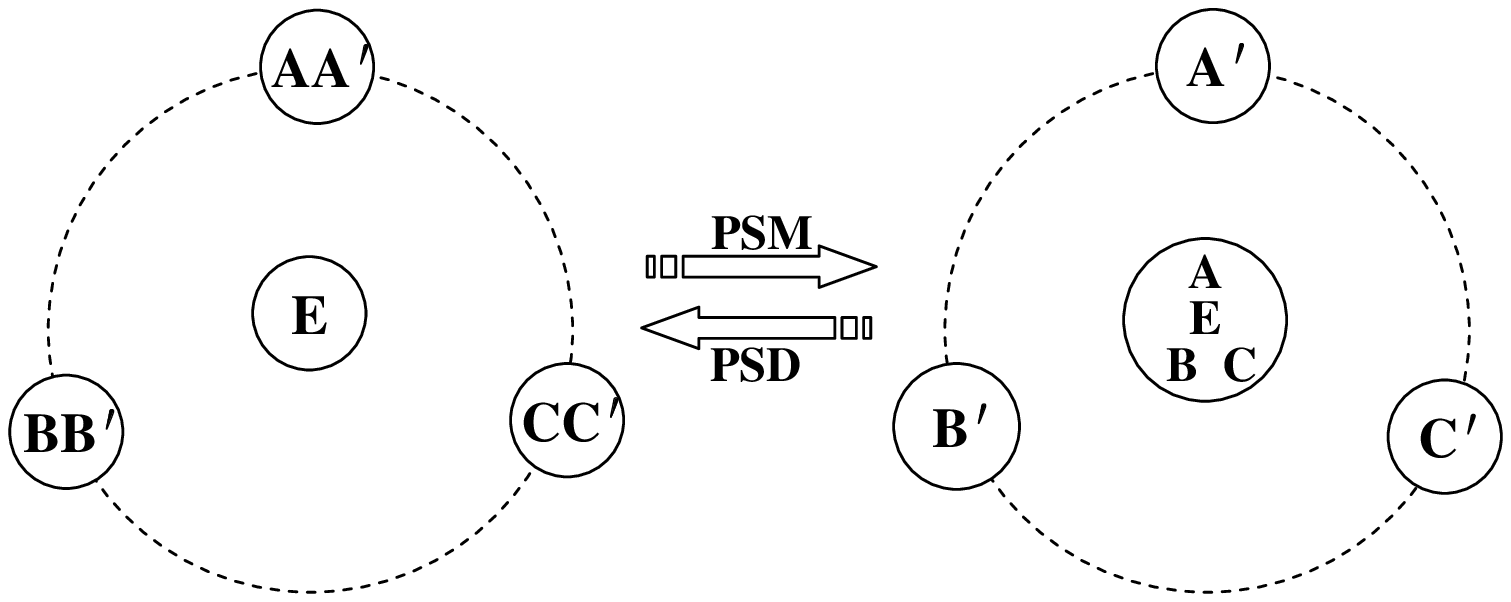}
\caption{Partial state merging and partial state distribution}
\end{figure}

One can check that $Q=\frac{1}{2}\{I(AA':BB')-I(A':B')\}$ is also the quantity that describes the flow of qubits out of the center in the process of partial state distribution (PSD) that is the reversed process of PSM. In FIG 4, we depict the two reversible processes for tripartite state. It is easy to see that the factor $1/2$ remains throughout calculating the flow of qubits. This gives an operational ground that the factor is $1/2$ even for multipartite entanglement. Notice that if only the monotonicity under LOCC is required, the factor can be taken for example $1/n$ for the n-partite case that is also reduced to the same formula for bipartite case. However it does not match the operational meaning.

In summary, we have constructed an additive entanglement measure---conditional entanglement of mutual information, and elaborated its operational meaning. The conclusions have been generalized to multipartite entanglement, with an additive and operational multipartite entanglement measure being provided for the first time.

{\it Acknowledgement} This work is supported by the RGC of Hong Kong (HKU 7051/06 and HKU 3/05C), the NSFC(10805043,10429401), the State Key Program for Basic Research of China (2006CB921800), the grant PBZ-MIN-008/P03/2003, and the EU IP SCALA. MH would like to thank P. Badziag for numerous discussions on quantum conditioning.


\begin{thebibliography}{99}
\bibitem{note}
A measure of correlations  is required at least to be nonincreasing under local operations, see \cite{HendersonVedral}.
\end{thebibliography}
\end{document}